 \documentclass[preprint]{aastex}

\newcommand{\naid}{\mbox{Na\,{\footnotesize I} {\sl D }}}
\newcommand{\cai}{Ca\,{\footnotesize I} }

\newcommand{\caiih}{Ca\,{\footnotesize II}~H }
\newcommand{\caiik}{Ca\,{\footnotesize II}~K }
\newcommand{\caiihk}{Ca\,{\footnotesize II}~H \&~K }
\newcommand{\mgi}{Mg\,{\footnotesize I} }

\newcommand{\ie}{{\em i.e.,} }
\newcommand{\eg}{{\em e.g.,} }

\newcommand{\lsi}{$\mathrel{\hbox{\rlap{\hbox{\lower2pt\hbox{$\sim$}}}\raise2pt\hbox{$<$}}}$}
\newcommand{\gsi}{$\mathrel{\hbox{\rlap{\hbox{\lower2pt\hbox{$\sim$}}}\raise2pt\hbox{$>$}}}$}

\shorttitle{Low Dispersion Digital Spectral Library}
\shortauthors{James}

\begin{document}


\title{A Digital Low Dispersion Spectral Library Covering the 
3500-7500\r{A} Region Using the SAAO Radcliffe 1.9m Telescope's 
Cassegrain Spectrograph}


\author{David J. James}

\affil{Cerro Tololo InterAmerican Observatory, Casilla 603, La Serena, Chile}
\email{djj@ctio.noao.edu}


\begin{abstract} 
We have created a digital spectral library, using low resolution optical 
spectra, of photometric and spectral standard stars. The data were acquired 
using the Cassegrain Spectrograph installed on the 1.9m Radcliffe telescope 
at the South African Astronomical Observatory. The library consists of 
optical wavelength ($\simeq$ 3500$-$7500\r{A}) spectra for main sequence 
and giant stars encompassing those most commonly observed in the Galaxy, 
namely the late-B, A-, F-, G-, K-, and early- to mid-M stars. We intend 
that our standard star spectra will be especially useful for spectral 
classification of stars in the field and Galactic clusters alike, and 
will have high pedagogic value when included into representative 
{\em Introductory Astronomy} or {\em Stellar Astronomy} curricula for undergraduate 
astronomy major and minor programs.

We exploit the spectral library in order to derive spectral types 
for seventy-six optically and X-ray selected members of the young 
open cluster {\sc ngc 6475}. Comparison of spectral-type, optical and 
infrared photometric data to theoretical colors derived from spectral 
type show that the reddening of the cluster is E$_{(B-V)}$ = 0.068 
$\pm$ 0.012  ($1\sigma$=0.058), a vector consistent with earlier 
surveys. Our analysis also highlights the utility of such spectra 
in rejecting cluster non-members, thereby allowing the creation of 
a clean sample of {\em bona fide} cluster members for follow-up 
science observations.  
\end{abstract}





\section{Introduction}

Exploiting low resolution optical spectra can be a powerful tool in establishing 
fundamental properties of celestial objects of astrophysical interest. This is 
facilitated by the fact that low resolution spectra are oftentimes easy 
to acquire and reduce, yield much higher count-rates per unit time than higher 
resolution spectra and the nature of the celestial object can readily be 
established without substantial post-reduction analysis. The scientific domain 
that such spectra engage is broad, and includes for instance, the determination 
of redshifts for extra-Galactic objects ({\em e.g.,} Folkes et al. 1999; 
Doroshkevich et al. 2004), producing reddening maps of young star formation 
regions ({\em e.g.,} James et al. 2006; James et al. 2013), identifying 
magnetically active solar-type stars ({\em e.g.,} Mart\'{\i}n et al. 1998; 
Jeffries et al. 1996) and classifying the nature of very red objects in 
near-IR photometric surveys ({\em eg.,} Kirkpatrick et al. 1999; Allende 
Prieto et al. 2004).

In today's golden era of large-scale, multi-filter, wide-field synoptic 
photometric surveys ({\em e.g.,} The Sloan Digital Sky Survey [{\sc sdss} - 
Castander 1998]; The Dark Energy Survey [{\sc des} - Abbott et al. 2005]; 
Vista variables in the Via Lactea [{\sc vvv} - Minniti et al. 2006]), and 
those to follow in the near-to-mid future ({\em e.g.,} The Large Synoptic 
Survey Telescope project [{\sc lsst}] - Ivezic et al. 2008), the r\^{o}le of 
single- and multi-object low resolution spectrographs is crucial, and 
is very well defined. In what essentially serves as a {\em classification engine} 
for these synoptic surveys, low resolution spectroscopy allows for 
quick and easy transient source identification, as well as for 
spectroscopic calibration of photometrically determined redshifts 
(so-called {\em photo-z} redshifts) for extra-Galactic sources. {\em Modulo}, 
a next-generation project for follow-up spectroscopy of synoptic survey 
transient sources has recently been proposed for possible future 
operation on the Cerro Tololo InterAmerican Observatory's [{\sc ctio}] 
Blanco 4m telescope, the Dark Energy Spectrometer [{\sc despec} -  
Abdalla et al. (2012)], and would serve as a natural segue from 
the now-active Dark Energy Survey.

Many previous compilations and libraries of stellar spectra in the optical 
and near-infrared exist,\footnote{{\em e.g.,} 
http://www.ucm.es/info/Astrof/invest/actividad/spectra.html} of differing 
qualities and spectral resolutions, however until recently, few were 
digitally available. The excellent compilations of Montes et al. (1997), 
Bochanski et al. (2006), and Lebzelter et al. (2012) are fine examples of 
online libraries now available, where each manuscript presents a fine 
historical overview of pre-internet spectral-type surveys.


The driving force behind our project, and this manuscript, is to provide 
a library of low resolution standard star spectra, in both print and digital 
format, for main sequence and giant stars covering the gamut of the most 
commonly observed types of stars in the Galaxy, namely the late-B, A-, 
F-, G-, K-, and early- to mid-M stars. Our standard star spectra will 
be especially useful for spectral classification of stars in the field 
and Galactic clusters alike, and are highly didactic when introduced, for 
instance, into the curricula of {\em Introductory Astronomy} type classes for 
astronomy major and minor undergraduates.

We herein present low-resolution spectra, totaling eighty-three (83) stars, 
for a sample of photometric and spectral-type standard stars, covering 
a large swathe (3500-7500\r{A}) of the optical wavelength region, 
approximately corresponding to bandpasses of the Johnson-Cousins 
B-, V- and R-filters (Bessell 1990). Details of the observations and 
data analysis are presented in \S~2, while a description of the spectral 
library is addressed in \S~3, together with commentary for a handful 
of special cases. In an appendix, we further use our spectral library 
to derive spectral types for photometrically and X-ray selected stars 
in the vicinity of the young open cluster \object[M7]{NGC 6475}, in 
order to establish a sample of {\bf bona fide} cluster members for 
follow-up study. Our analysis highlights the utility of such spectra 
in rejecting cluster non-members and establishing reddening vectors, 
leading to a reddening map, for genuine cluster members.

\section{Observations and Data Reduction}\label{Obs-spec}

Low resolution, single-order spectroscopy has been obtained for eighty 
three bright standard stars, having well-established extant spectral 
types and luminosities, during the period 29 June - 12 July 1999, using 
the 1.9~m Radcliffe telescope at the South African Astronomical 
Observatory [{\sc saao}], Sutherland, South Africa. The {\sc saao} 
observations were performed using the Cassegrain spectrograph, a 
300 line mm$^{-1}$ grating blazed at 4600\r{A} ({\sc saao} \#7) 
used in first order, a 1.8-arcsecond slit width and a 15-micron 
pixel, $266\times1798$ {\sc sit}e {\sc ccd} in 2x1 binning mode, as 
the detector. This set-up yielded a three-pixel spectral resolution 
of $\sim 7$\r{A} at 5500\r{A}, R$\simeq$800, and a spectral range of 
$\sim 3550-7550$\r{A}. A full mass and luminosity range of 
photometric and spectral standard stars (B2$-$M6; luminosities I$-$V; 
Garcia 1989; Keenan \& McNeil 1989; Keenan \& Pitts 1980; 
Kilkenny \& Cousins 1995; Menzies \& Laing 1988; Morgan \& Keenan 
1973; Vogt, Geisse \& Rojas 1981) were observed with the same 
instrumental set-up as used for the {\sc ngc 6475} target stars. An 
observing log of our {\sc saao} spectroscopic observations of 
standard stars is presented in Table~\ref{OBStable}.

Standard data reduction techniques (de-biasing and overscan 
subtraction) were employed on raw {\sc ccd} frames using routines 
in the Starlink\footnote{http://starlink.jach.hawaii.edu/starlink} 
data reduction algorithm suite. Flat-fielding of the bias-subtracted 
spectra was achieved using a balance frame created from a median series 
of quartz lamp exposures. Spectra were extracted using an optimal 
extraction algorithm (Horne 1986), with local background sky 
subtraction. Wavelength calibration was achieved by reference to 
ThAr arc lamp spectra bracketing target spectra. Wavelength 
calibrated, extracted spectra were trimmed to a spectral window 
of $3640-7500$\r{A}, and normalized using {\em spline3} polynomials 
fits, ranging from 5$^{th}$ to 15$^{th}$ order depending on 
spectral type and luminosity. 

While we do not feel that the entire spectral library is suitable 
for flux calibration purposes due to the relatively narrow slit used, 
the inhomogeneity of sky conditions over the course of the observing 
run, and the need for neutral density filters for the brighter objects 
in the sample, we provide flux calibration for sample spectra acquired 
during the UT19990702 observing night. Flux calibration using 
dedicated Starlink software routines was performed on the extracted 
spectra, re-binned to a logarithmic wavelength scale (\ie constant 
velocity steps) and converted into flux units of counts per second. 
These spectra were subsequently corrected for atmospheric extinction 
at the Sutherland site of the {\sc saao} using recently updated 
relations\footnote{http://www.salt.ac.za/observingtools/pipt/pipt\_salticam\_simulator/0.5/doc/api/src-html/za/ac/salt/pipt/salticam/spectrum/Atmosphere.html} 
fit to the Spencer Jones (1980) study. Extinction corrected count rates were 
converted to flux (in Janskys) by comparison to a spectrum of the 
spectrophotometric flux standard star LTT 7379 and its tabulated fluxes 
from Stone \& Baldwin (1983). Finally, fluxed spectra were transformed 
to units of ergs s$^{-1}$ cm$^{-2}$ \r{A}$^{-1}$.


\section{The Spectral Library Sample}

Our sample of photometric and spectral-type standard stars was 
selected from several disparate sources, where the primary 
criteria for observation were for each star to have a known and 
well-established spectral-type as well as stable optical photometry. 
Each star was chosen from sets of E- and F-region photometric standard 
stars (Vogt et al. 1981, Menzies \& Laing 1988, Kilkenny \& 
Cousins 1995) and the Garcia (1989) collation of MK standard 
stars. A full listing of spectral type, special physical properties 
and optical UBV(RI)c photometry for each star in the spectral 
library is presented in Table~\ref{SpTytable}, where data 
references for each spectral-type are shown in columns~4 \& 10. 
For completeness, we also include a table of near-infrared {\sc 2mass jhk} 
photometry for each star in Table~\ref{2MASStable}, although 
we note that there are problematic error codes for about two-thirds 
of the sample, mostly associated with their extreme brightness for 
that survey.


{\em A priori}, none of the stars was checked for signatures of magnetic 
activity, which for a set of photometric-invariant field stars in 
the Galaxy, should reveal that most objects are chromospherically and 
coronally inactive. For each star in Table~\ref{SpTytable}, physical 
property comments (in column 2) show that this supposition is generally 
correct, whereby only thirteen stars have {\sc rosat} All Sky Survey 
[{\sc rass}] detections. Moreover, very few stars, three late-M dwarfs 
and one Be star, show evidence of H$\alpha$ in emission, which for 
later-type stars is typically associated with magnetic field-induced 
chromospheric emission. A handful of stars are multiples (visual or 
spectroscopic doubles, and one triple), which for the majority of 
them should not affect their spectral type determination (for instance, 
we were careful to avoid having visual double companions on the 
spectrograph slit together with the primary target). Six stars in 
the library have been shown, or are postulated, to exhibit signatures 
of extra-solar planets orbiting them, which if confirmed, should 
have no, or very little, effect on their spectral type determinations 
or their optical spectra.

\begin{figure}
\vspace*{-15mm}\includegraphics[angle=0,scale=.72]{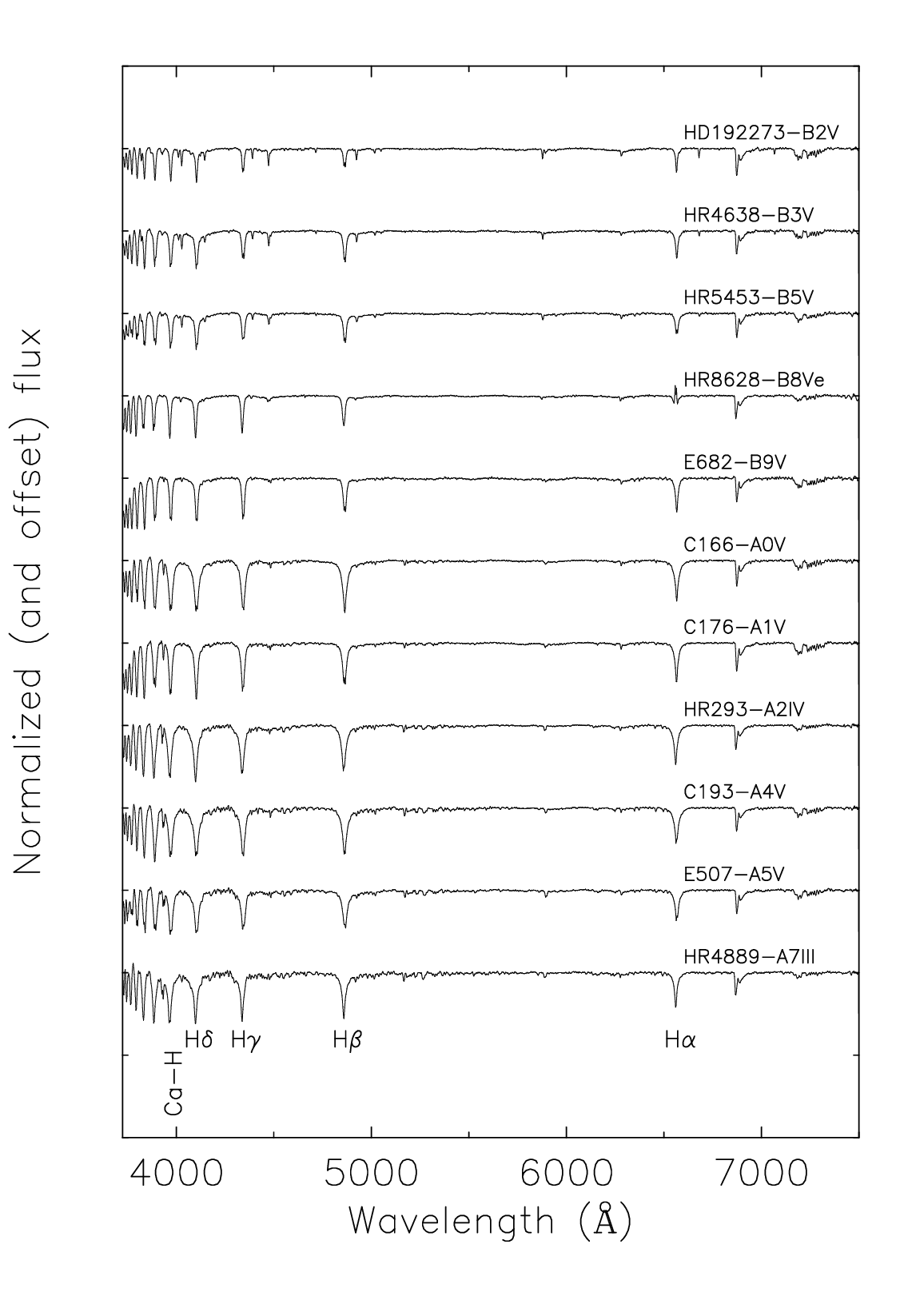}
\caption{Normalized (and offset) spectra for a subsample of objects detailed in 
Table~\ref{SpTytable}, obtained using the instrumental set-up described in 
\S~\ref{Obs-spec}, are presented for stars in the spectral range B2V $-$ A7 III. 
Identification of major spectral features are annotated at the bottom of the 
composite spectra.}
\label{Figure1}
\end{figure}

\begin{figure}
\vspace*{-15mm}\includegraphics[angle=0,scale=.72]{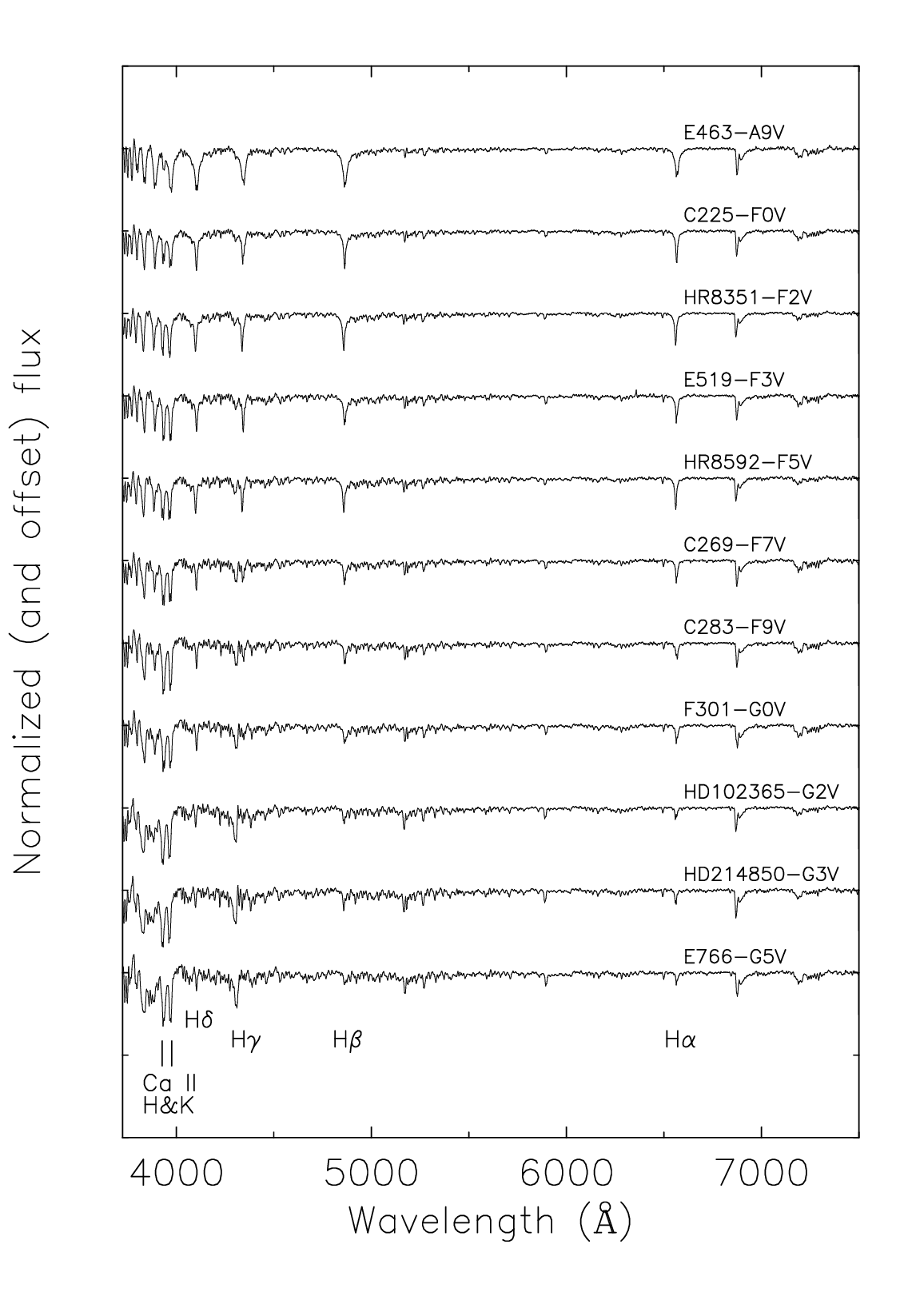}
\caption{Same as Figure~\ref{Figure1}, for stars in the spectral 
range A9V $-$ G5V.}
\label{Figure2}
\end{figure}

\begin{figure}
\vspace*{-15mm}\includegraphics[angle=0,scale=.72]{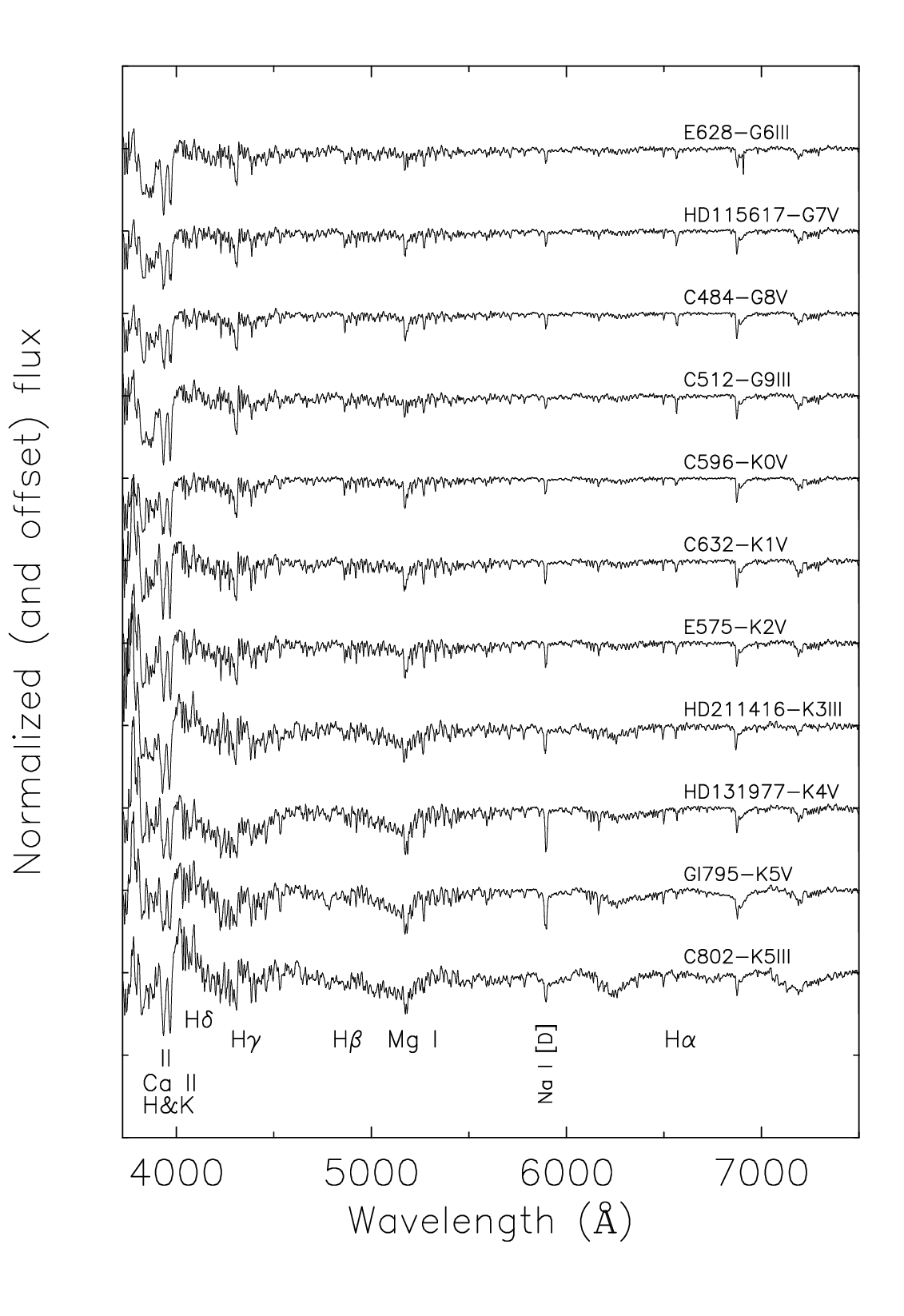}
\caption{Same as Figure~\ref{Figure1}, for stars in the spectral 
range G6 III $-$ K5 III.}
\label{Figure3}
\end{figure}

\begin{figure}
\vspace*{-15mm}\includegraphics[angle=0,scale=.72]{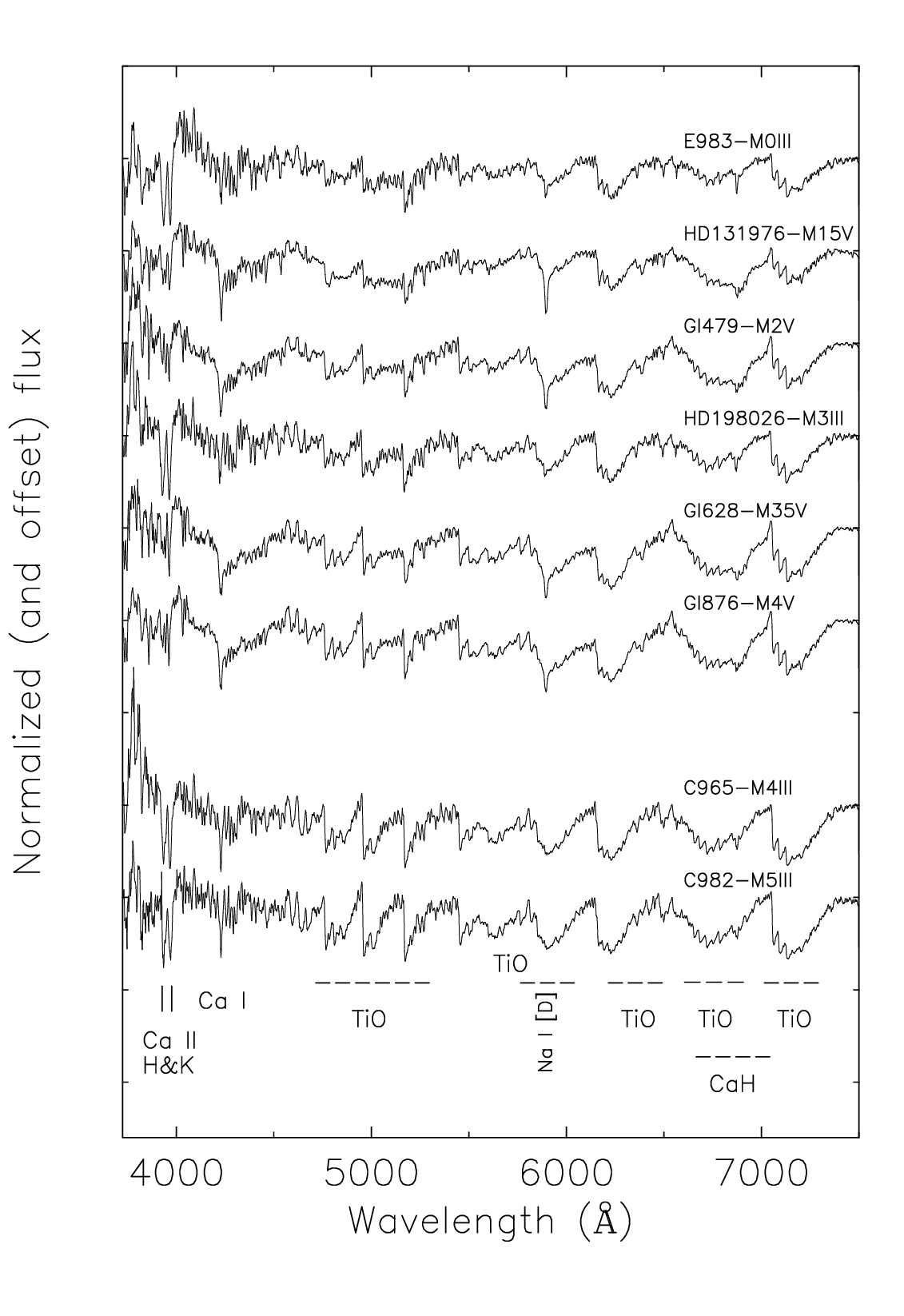}
\caption{Same as Figure~\ref{Figure1}, for stars in the spectral 
range M0 III $-$ M5 III.}
\label{Figure4}
\end{figure}

Example rectified {\sc saao} low-resolution spectra are shown for 
representative stars of the spectral library in Figures~\ref{Figure1}, 
~\ref{Figure2}, ~\ref{Figure3} \& \ref{Figure4}. We issue a {\em caveat lector} 
when using rectified spectra for the later-type stars because of 
difficulties in placement of the continuum in spectral regions heavily 
affected by molecular bandhead absorption. The un-normalized {\em trimmed} 
spectra for these stars, in counts or absolute flux units, should be 
preferentially employed in analyses. For the earlier type stars (\lsi F0V), 
the first five or six lines of the hydrogen Balmer series and the 
\caiih (3968 \r{A}) line are very prominent, with few metallic lines 
evident. For the mid-mass range, F0V $\rightarrow$ K0V, the low 
resolution spectra change appearance appreciably with the growth in 
strength of the \caiik (3934 \r{A}) line, the \mgi triplet lines 
(5167, 5173 \& 5184 \r{A}) and the \naid doublet at 5890 \& 5896 
\r{A}. For the lowest-mass, late-type stars (\gsi K0V), the 
Ca\,{\footnotesize II}~H \&~K, \mgi triplet and \naid lines 
still feature prominently, although the appearance of TiO and CaH 
molecular bandheads become ever more pronounced toward later 
and later types.

\subsection{Homogeneity of Spectral Classifications}\label{homoSpTy}

We are most grateful to the manuscript referee for urging us to consider 
the homogeneity of our spectral library, and to provide appropriate 
discussion for the reader. Because the spectral library comprises spectral 
types derived from several disparate reference materials in the literature, 
it is essential and didactic to consider the absolute value and 
inter-comparability of each spectral type provided in Table~\ref{SpTytable}.

In the first instance, in order to better understand the homogeneity 
of our spectral system we cross-correlate in Fourier space every star 
in the library with every other one. We subsequently search for the 
corresponding matching spectrum which yields the highest\footnote{For a 
small handful of cross-correlation functions, we choose the template 
star that yielded the second highest peak (in all cases smaller by 
only $1\%$) because it returned a far higher Tonry \& Davis (1979) 
{\em R} number.} cross correlation function height -- where 1.0 is the 
maximum. Results of this process are detailed in Table~\ref{SAAO-selfCCF}, 
where we list the corresponding template that best matches with each 
target spectrum in the library, and note differences in both spectral 
type and luminosity class. The mean difference in derived spectral 
types is -0.1 in class ($1\sigma$=1.75), which indicates that the 
zero-point of the spectral type system is robust, but has a scatter 
of about 2 sub-classes. Curiously, there are two stars, C~112 and C~512, 
which have substantially mis-matched spectra at the 4 \& 5 sub-class 
level, respectively, although we note that C~112 is a known 
variable star. 

\begin{figure}
\vspace*{-15mm}\includegraphics[angle=0,scale=.72]{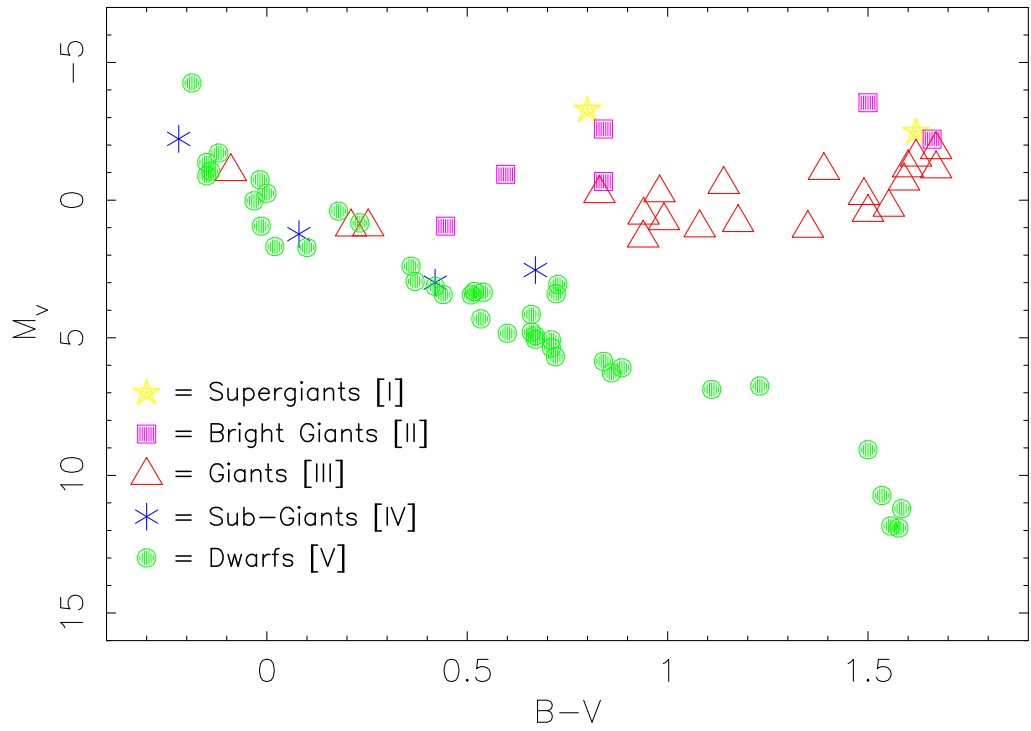}
\caption{Hertzsprung Russell diagram for stars in the spectral library, 
without interstellar extinction corrections, triaged into luminosity 
class. Distances are based are targets' {\sc hipparcos} trigonometrical 
parallaxes detailed in Table~\ref{HIPtable}.}
\label{Figure5}
\end{figure}

In the second, we construct an Hertzsprung-Russell diagram 
[{\sc hrd}] in M$_{v}$, B-V space (without reddening corrections) using 
{\sc hipparcos} astrometric data for each star in the library, which 
are detailed in Table~\ref{HIPtable}. Plotted in Figure~\ref{Figure5}, 
the library stars show the textbook, canonical form of an {\sc hrd}, 
and with the exception of three hotter, blue giants, the overall 
morphology of the data sample matches expectation. The loci of the 
giant-branch and main sequence are very well-defined, with 
appropriate positions for the bluer (B-V$>$0.3) bright giants and 
supergiants $-$ although the upper and lower boundaries of each 
are not as tightly-defined and possibly overlap. Positions of the 
few sub-giants in our sample are not as quite as well-defined as 
one would expect, as three out of the four stars with {\sc hipparcos} 
parallaxes appear more closely aligned with the main sequence 
locus.

Interestingly, in about $25\%$ of cases, our cross correlation analysis 
reveals a shift in luminosity class. We attribute these shifts to one 
of four plausible causes, and urge caution in over-reliance of absolute 
luminosity classification using the library spectra alone: {\bf (i)} a lack 
of luminosity class resolution in the library, where not enough stars of 
a given spectral type and luminosity class were observed to sufficiently 
resolve each star by the correlation method $-$ this effect is especially 
true for the earliest type stars; {\bf (ii)} spectra of intrinsically brighter 
supergiants, giants or sub-giants mimic the spectral properties of slightly 
fainter temperature giants, sub-giants or dwarfs, or vice versa; {\bf (iii)} the 
stars are variable; {\bf (iv)} mis-identifications in the original, published 
spectral classifications.

\begin{figure}
\hspace*{-15mm}\includegraphics[angle=0,scale=.72]{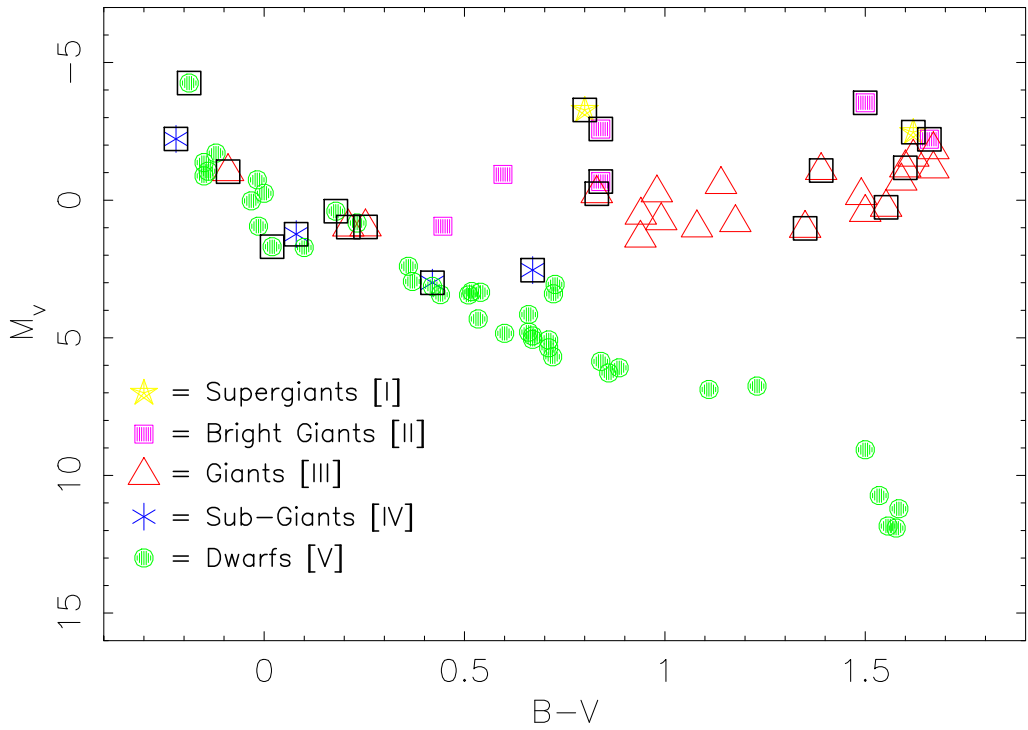}
\caption{Reproduction of Figure~\ref{Figure5} to highlight those objects 
(with square symbols) demonstrating a luminosity class change in the cross 
correlation analysis described in \S~\ref{homoSpTy}.}
\label{Figure6}
\end{figure}

While the aforementioned point {\bf (i)} may indeed be affecting the cross 
correlation results, we can use the position of stars in the {\sc hrd} to 
test effects {\bf (ii)}, {\bf (iii)} \& {\bf (iv)}. We should note however that 
the physical interpretation of one or all of the last three effects is 
degenerate in the sense that one or all of them can mimic displacement 
in the M$_{v}$, B-V plane. In order to guide the eye in Figure~\ref{Figure6}, 
we re-plot Figure~\ref{Figure5} highlighting instead those stars which 
show luminosity class changes ({\em c.f.,} Table~\ref{SAAO-selfCCF}) in 
the cross correlation analysis. We can divide the problematic 
luminosity stars into two broad bins, those with B-V$<$0.60, and those 
redward of B-V$=$0.60, while understanding that some subset of the 
highlighted stars may be variable, the following points are evident:

\begin{itemize}
\item The sub-giants are not well-defined, and one could argue $-$ in agreement 
with the cross correlation results in Table~\ref{SAAO-selfCCF} $-$ that the 
three class-IV stars blueward of B-V=0.6 could be re-classified as dwarf [V] 
stars based on our epoch-1999 spectra. The position of the reddest sub-giant is 
appropriate for its class.

\item The three bluest giant [III] stars are inconsistent with the loci of 
various flavors of well-established giant branches, and based upon our 
epoch-1999 spectra, are in fact representative of dwarf objects. We note that 
two of them, C 112 and HR 4889, have luminosity classes reported in the 
broader literature of III, IV and V.

\item
The four dwarf stars blueward of B-V=0.60, with peak-heights in the 
cross correlation analyses that showed a luminosity class change 
(to either IV or III), show no evidence of {\sc hrd} displacement from 
their expected dwarf positions.

\item
For the remaining stars redward of B-V$=$0.60, the picture is hazier. 
In Table~\ref{SAAO-selfCCF}, there are several giants and bright giants 
that are identified as either changes up to class-II or down to class-III. 
In the {\sc hrd}, the giants and bright giants are appropriately located for 
the most part, and one cannot easily distinguish between the brighter 
class-III objects or the fainter class-II objects where the populations 
overlap. The location of the two supergiants in the {\sc hrd} is as one 
would expect, although a lack of interstellar reddening/extinction 
correction may be acting to make them appear slightly under-luminous. 

\item
Finally, we note that one late-type giant in the library, HD 114873, 
is classified as a dwarf star of the same spectral type in the cross 
correlation analysis. Its {\sc hrd} position (M$_{v}$=1.03, B-V=1.35) 
is exactly where one would expect a late-type giant to be found, and 
we can therefore rule out its dwarf status based upon our epoch-1999 
spectra.
\end{itemize}

\section{Digital Spectral Library}

In order to make our spectral library widely available to the wider 
astronomical community, especially in digital format, all data products 
(trimmed- and continuum-fitted spectra, {\sc ascii} tables of trimmed 
spectra (and flux-calibrated spectra of targets for the observing 
night 02-July-1999)) for the entire catalog are available for web 
download\footnote{http://www.saao.ac.za/science/facilities/instruments/grating-spectrograph-with-site-ccd/spccd-spectral-type-library/} 
or via ftp\footnote{ftp.ctio.noao.edu/pub/djj/{\sc saao}-SpectralType-Library/}. A description of 
the data structure and format is included in as a {\em ReadMe} file in each data repository.

\acknowledgments

The author acknowledges the data analysis facilities provided by the 
Starlink Project which is run by {\sc cclrc} on behalf of {\sc pparc}. In 
addition, the following Starlink packages have been used: {\sc echomop} 
and {\sc figaro}. This research has also made use of the {\sc simbad} 
database, operated at Centre de Donn\'{e}es Astronomiques de Strasbourg, 
Strasbourg, France, as well as the facilities of the Canadian 
Astronomy Data Centre operated by the National Research Council 
of Canada with the support of the Canadian Space Agency. This 
publication also makes use of data products from the Two Micron All 
Sky Survey, which is a joint project of the University of 
Massachusetts and the Infrared Processing and Analysis 
Center/California Institute of Technology, funded by the National 
Aeronautics and Space Administration and the National Science 
Foundation. Some assistance in the spectral typing of M7 stars 
was provided by Nick Dunstone, and is gratefully acknowledged. 
We would also like to offer heartfelt thanks for fruitful discussions 
with Eric Mamajek (U. Rochester), especially for sharing his historical 
spectral type notes. Finally, we are particularly thankful and grateful 
to our manuscript Referee who provided a rapid, fair and well-balanced 
referee's report for our project.



{\it Facilities:} \facility{South African Astronomical 
Observatory (Radcliffe 1.9m telescope)}.




\clearpage



\appendix
\section{Spectral types of stars in M7}

The Genesis of this project, nearly 15-years ago, was the desire to 
obtain low resolution spectra for photometrically-selected members of 
the young open cluster, {\sc ngc 6475} (M7), the main astrophysical target 
of interest for the author's doctoral thesis. Observing Galactic open 
clusters is a noble pursuit because they are unique astrophysical 
observatories fundamental to studying stellar evolution. Their great 
utility arises because their ages can be relatively accurately be 
determined by isochrone fitting of cluster color-magnitude diagrams 
[{\sc cmd}s $-$ ({\em e.g.,} Sandage 1958; Demarque \& Larson 1964; 
Naylor 2009; Cargile \& James 2010)], determination of the lithium 
burning-limit age ({\em e.g.,} Basri et al. 1996; Cargile, James \& 
Jeffries 2010) or by fitting gyrochronology models to cluster 
photometric period-color distributions  ({\em e.g.,} Barnes 2003, 2007, 
2010; James et al. 2010). 

M7, also known as Ptolemy's cluster, is a fairly well-populated southern 
hemisphere object in the constellation of Scorpius, situated about 
$4^{\circ}$ south of the Galactic plane [$17^{h}53^{m}, -34^{\circ}52.8^{'},$ J2000]. 
Initial investigations of M7 were primarily concerned with obtaining photometry 
for the brighter, [V$\leq 11$], higher mass members of the cluster. Johnson 
{\sc ubv} photoelectric and photographic photometry were obtained down to spectral 
type $\sim$F5 by Koelbloed (1959) and Hoag et al. (1961). Approximately 15 years 
later, these photometric data were used to facilitate spectral-type determinations 
within the cluster (Abt 1975), in order to investigate interstellar reddening 
in the direction of the cluster and evaluate its distance (251 pc; Snowden 1976). 
Thereafter, a small sample of high mass cluster members were observed by various 
authors to study binarity, variability and conduct radial velocity surveys amongst 
B and A stars of the cluster (\eg Leung \& Schneider 1975; Engberg 1983; Gieseking 1985).

The most recent robust age, distance and reddening estimates for the cluster 
are given by Meynet, Mermilliod \& Maeder (1993) using isochrone fitting. 
Employing solar-metallicity isochrones, incorporating convective overshoot and 
improved opacity tables, they fit the high mass members in the color-magnitude 
diagram, and obtained a distance of $\sim$240~pc, a reddening of E$_{B-V}$=0.06, 
and an age of 224~Myr (the age derived by Mermilliod 1981, and consistent with 
the trigonometric parallax distance [270 pc] derived by van Leeuwen 2009). 
James \& Jeffries (1997 - [JJ97]) performed the first extensive, high-resolution 
spectroscopic observations of the cluster's solar-type stars in their study 
of radial and rotation velocities, chromospheric magnetic activity and lithium 
abundances of X-ray selected cluster members. They exploited their \'{e}chelle 
spectra to derive a slightly super-solar metallicity for the cluster of 
[Fe/H]$=+0.110 \pm 0.034$. Sestito et al. (2003) performed a second, high-resolution 
spectroscopic study of M7, specifically focused on lithium abundance evolution of 
its solar-type stars, and with a cluster metallicity determination of 
[Fe/H]$=+0.14 \pm 0.06$, in agreement with the JJ97 result. However, a recent 
metallicity determination of seven stars in the cluster (Villanova et al. 2009), 
based on European Southern Observatory archival spectra, yields a more 
solar-like metallicity of [Fe/H]=$+0.03 \pm 0.02$. We note however that they 
included two dwarf B-stars and two late-giants in their analysis, which for 
cluster abundance work is somewhat unusual and can lead to abundance determination 
irregularities (for the lack of metallic lines and possible dredge-up of 
processed material, respectively).

Performing a low resolution spectroscopic survey of open clusters like 
{\sc ngc 6475} represents a quick-and-easy, resource-cheap and scientifically 
powerful method of identifying {\em bona fide} cluster members for follow-up 
study through rejection of foreground dwarf stars and background giants. 
This methodology was particularly useful for X-ray pointings of crowded 
fields in the {\sc rosat} era, where positional error circles for 
{\sc pspc} observations were large (typically 25-30$^{"}$ at 1~keV - Briel 
\& Pfeffermann, 1986), and remained problematic for the most crowded 
fields even with the {\sc hri} ($\simeq 5^{"}$ - David et al. 1996). Because 
{\sc ngc 6475} lies only four-degrees south of the Galactic plane, such large 
spatial error circles meant that identifying optical counterparts in X-ray 
error circles that were likely to be true cluster member was fraught with 
uncertainty. This in turn affects one's ability to establish cluster 
membership using only a photometric catalog, mostly due to the large 
number of non-member, field star interlopers contaminating the cluster's 
photometric sample. 

Having observed M7 stars during the same {\sc saao} observing run 
used to obtained spectra for our spectral library, we have been able 
to perform a spectral type analysis of several photometrically and 
X-ray selected members of {\sc ngc 6475}. All spectral type determinations 
were made by visual comparison to our new library, with special 
attention being given to the \caiihk lines, \cai lines at 4226, 
6122, 6162\r{A}, \naid lines at 5890 \& 5896\r{A}, and the first 
four Balmer series lines at 6562.8, 4861.3, 4340.4 \& 4101.7\r{A} 
(H$\alpha \rightarrow \delta$), as well as the TiO bandheads such 
as those at 4760, 5167 \& 5448\r{A}, 6700\r{A} and 7200\r{A} for 
the cooler targets. We have shown that spectral types are reliable 
to $\pm$ two sub-divisions.

Bright, hot star targets were selected from the photographic and 
photoelectric sample by Koelbloed (1959- [K59]), whereas the {\sc BVI}c 
{\sc ccd} survey of the cluster by Prosser et al. (1995 - [P95]) was 
employed for the fainter, lower-mass stars. Details of the M7 targets 
are presented in Table~\ref{M7table1}, with column~2 reserved for 
cross-identifications with the JJ97 spectroscopic study. Astrometry 
and near infrared magnitudes (obtained from {\sc 2mass} sources) are 
provided for each M7 object in columns 3, 4, and 8-10 respectively. 
Optical photometric data, taken from K59 and P95, for V, B-V and V-Ic 
(where available) are listed in columns 5-7. 

\begin{figure}
\hspace*{-15mm}\includegraphics[angle=-90,scale=.64]{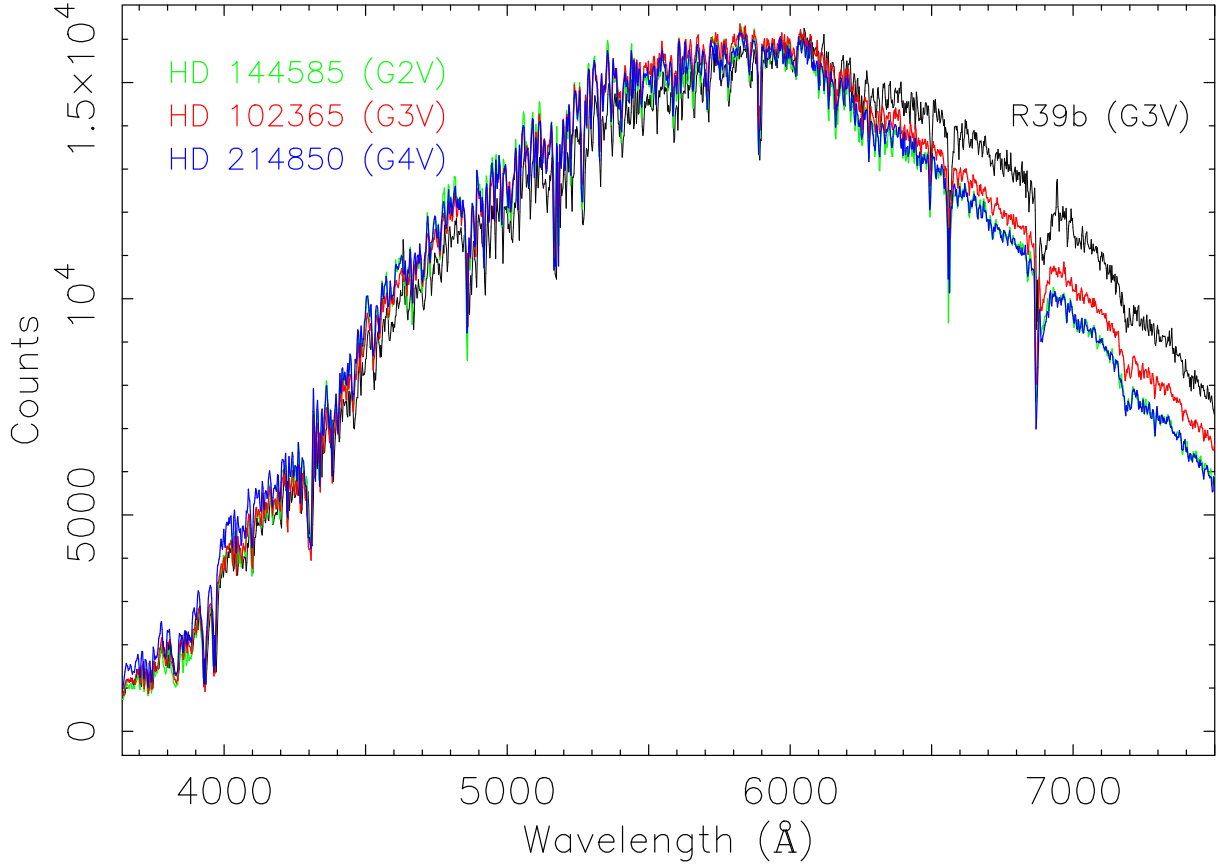}
\caption{{\sc saao} low resolution spectra are presented for the 
M7 cluster member R39b (see Table~\ref{M7table1}), together with 
comparison reference spectra from our spectral library, from which 
we infer a spectral type of G3V for R39b. The reference spectra 
are scaled to the flux of R39b at 6000 \r{A}. We note the poor 
spectral match in the red (\gsi 6200 \r{A}), which may be caused 
by additional red flux from an unseen, low-mass secondary companion 
(JJ97 find this star to be a single-lined spectroscopic binary).}
\label{Figure7}
\end{figure}

Our derived spectral types for M7 stars are detailed in 
Table~\ref{M7table2}, with an exemplar of the spectral typing process 
shown in Figure~\ref{Figure7}. We note that H$\alpha$ and \caiihk 
move into overt emission (emission above the local continuum level) 
at spectral type K3V/K4V. For each M7 star, where the photometric data 
allow, we also calculate reddening vectors E$_{B-V}$, E$_{V-Ic}$, 
E$_{J-H}$ and E$_{H-K}$ by comparing observed photometric colors to 
Kenyon \& Hartmann (1995 - [KH95]) theoretical colors derived from 
spectral type. In establishing reddening vectors, we only select stars 
brighter than V=12 magnitudes because of deep reservations concerning 
source confusion and photometric quality in the P95 survey. This is 
because they chose to use aperture photometry instead of point-spread 
function [{\sc psf}] fitting photometry to measure stellar flux on 
their {\sc ccd} images, which from our experience is most unreliable 
for V$>$12 sources in open cluster fields so close to the Galactic 
plane $-$ due to contaminating flux arising from near-neighbors.

For all M7 stars listed in Table~\ref{M7table2}, having V$\leq$12, we 
find a mean E$_{(B-V)}$ = 0.068 $\pm$ 0.012  ($1\sigma$=0.058, n=24). 
This value is consistent with previous determinations of B-V reddening 
in the cluster by K59 and Snowden (1976), and matches well with that 
determined by Maitzen \& Floquet (1981). In terms of V-Ic color, we 
find a mean E$_{(V-Ic)}$ = 0.049 $\pm$ 0.018 (1$\sigma$=0.069, n=14), 
which is nearly $50\%$ smaller than one would expect assuming a 
reddening law of E$_{(V-Ic)}$ = 1.30 $\times$ E$_{(B-V)}$ 
(Ram{\'{\i}}rez \& Mel\'{e}ndez 2005), although there is agreement 
within 1$\sigma$. In the near infrared, for the J-H and H-K colors, 
we find a mean E$_{(J-H)}$ = 0.059 $\pm$ 0.009 (1$\sigma$=0.053, n=32) 
and a mean E$_{(H-K)}$ = 0.008 $\pm$ 0.006 (1$\sigma$=0.034, n=32). Both 
of the infrared values lie within the 1$\sigma$ error range of expected 
E$_{(J-H)}$ and E$_{(H-K)}$ reddening values based on the Ram{\'{\i}}rez 
\& Mel\'{e}ndez E$_{(B-V)}$ reddening laws. 

\begin{figure}
\hspace*{-15mm}\includegraphics[angle=-90,scale=.64]{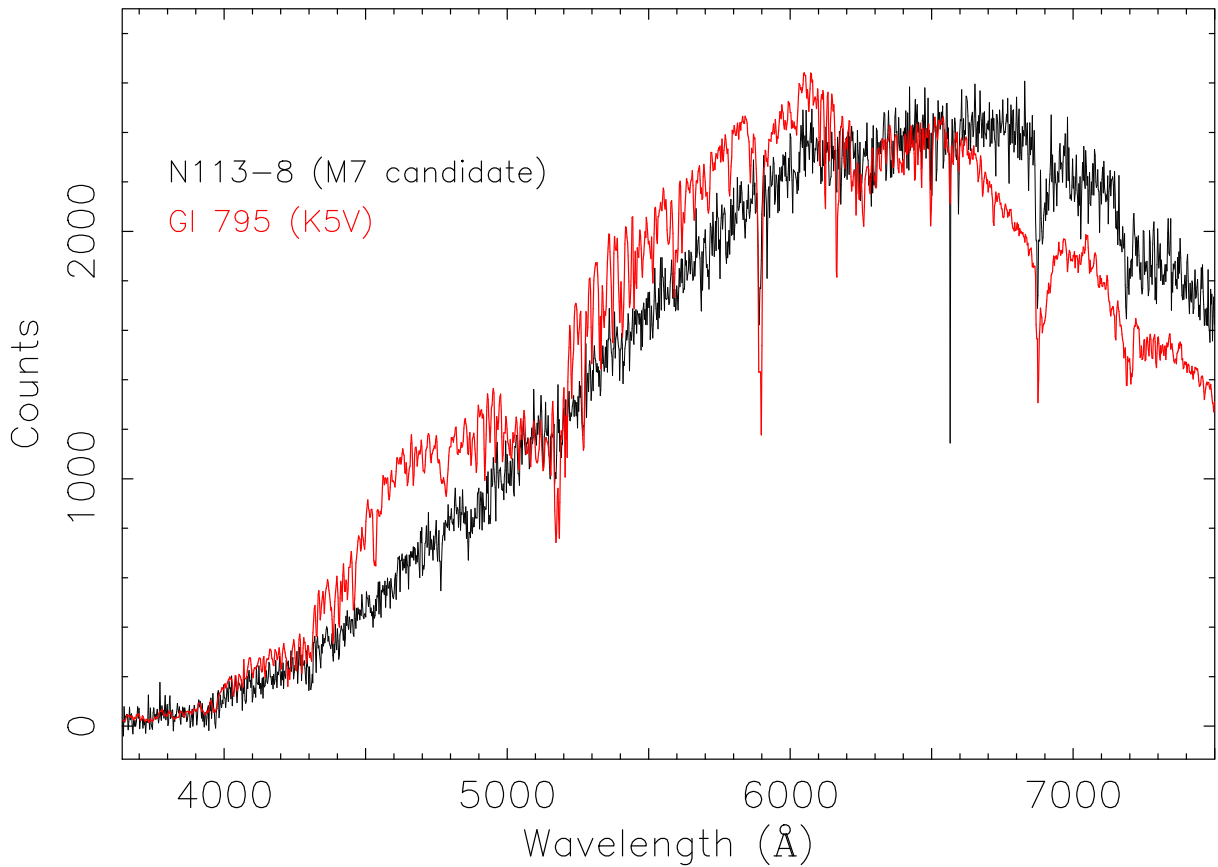}
\caption{An {\sc saao} low resolution spectrum is presented for M7 
candidate member N113-8 (candidacy based on author's doctoral thesis 
research), with {\sc ra}(2000): 17 54 18.6, {\sc dec}(2000): -35 25 38.0,  
V=15.33 and B-V=1.32. Also plotted, from the spectral library, is a 
spectrum of the K5V star Gl 795. The reference spectrum is scaled to 
the flux at H$\alpha$ (6563\r{A}).}
\label{Figure8}
\end{figure}

Comparison of low resolution spectra for candidate members of the 
cluster to our spectral library spectra is a powerful and efficient 
method of weeding out field-star interlopers contaminating the cluster 
sample. Such objects must be cluster non-members either lying in-front 
of, or behind, the cluster. An example of this methodology is shown in 
Figure~\ref{Figure8}, and shows how easily one can discover cluster 
non-members in the sample. In this case, we plot the {\sc saao} spectrum 
of an M7 candidate member of the cluster, {\em N113-8}, selected from the 
author's doctoral thesis research, with a dereddened (E$_{B-V}$=0.07) 
B-V$_{o}$=1.25. For this color, the K95 color-spectral type 
relationships show that this object should be spectral type $\sim$ 
K6V. In Figure~\ref{Figure8}, we also plot the library spectrum of 
the K5V star, Gl 795, with its flux matched to that of {\em N113-8} at 
H$\alpha$. There is a clear mis-match between the K5V standard and 
the M7 star, and strong evidence of suppressed blue flux, indicative 
of high interstellar reddening. Furthermore, the morphology of the 
M7 stellar spectrum, blueward of $\simeq$ 6000\r{A}, is particularly 
flat and featureless, indicative of a much hotter star than the 
expected K5V one. We posit that M7 candidate member {\em N113-8} 
is in fact a non-member, background reddened giant, of much 
earlier spectral type than would be expected for its de-reddened 
B-V color.


%
%



\clearpage

\clearpage

\clearpage

\clearpage

\clearpage

\clearpage

\clearpage

\clearpage

\clearpage

\clearpage

%
%

\clearpage

\begin{deluxetable}{llcrccccl}
\tablewidth{0pt}
\tabletypesize{\scriptsize}
\tablecaption{Properties and observation log for stars of known spectral type and 
luminosity observed during this program}
\tablehead{
\colhead{Star\tablenotemark{a}} & \colhead{Alt ID\tablenotemark{b}}  & \colhead{RA(2000)} & 
\colhead{DEC(2000)} & \colhead{Obs}    & \colhead{UT}  & 
\colhead{Exp}                   & \colhead{HJD}     & \colhead{Comments\tablenotemark{d}}  \\
\colhead{}                      & \colhead{}        & \colhead{[2MASS]\tablenotemark{c}}  
& \colhead{[2MASS]\tablenotemark{c}}   & \colhead{Date}   & \colhead{}    & 
\colhead{Time [s]}              & \colhead{[days]}  & \colhead{} }
\startdata
C 357      & HD 1835            & 00 22 51.74  & -12 12 34.08  &  1999-07-03  &    04:21 &  12   &   2451362.682   &                 \\
C 903      & HD 1879            & 00 23 04.39  & -15 56 33.58  &  1999-07-04  &    02:45 &  5    &   2451363.615   &                 \\
HR 293     & HD 6178            & 01 02 26.44  & -31 33 07.05  &  1999-06-30  &    03:31 &  26   &   2451359.647   &    thin cloud   \\
C 512      & HD 9362            & 01 31 15.11  & -49 04 21.46  &  1999-07-09  &    04:45 &  4    &   2451368.699   &    1.5  ND      \\ 
E 109      & HD 9403            & 01 31 32.65  & -43 50 47.70  &  1999-07-04  &    03:32 &  35   &   2451363.648   &                 \\
C 283      & HD 10647           & 01 42 29.32  & -53 44 27.11  &  1999-07-03  &    04:07 &  3    &   2451362.672   &                 \\
C 632      & HD 10476           & 01 42 29.78  &  20 16 07.36  &  1999-07-03  &    05:04 &  2    &   2451362.709   &                 \\
HD 10700   & HR 509             & 01 44 04.02  & -15 56 14.19  &  1999-06-30  &    02:45 &  4    &   2451359.613   &    thin cloud   \\
\enddata
\tablenotetext{a}{C$nnn$ - Garcia (1989) identifiers; E$nnn$ - E-region photometric standard 
stars (Vogt, Geisse \& Rojas 1981; Kilkenny \& Cousins 1995); F$nnn$ - F-region photometric 
standard stars (Menzies \& Laing 1988)}
\tablenotetext{b}{Alternative identifications, which include the wide-spread monikers: 
{\em GJ - Gliese \& Jahreiss}, {\em HD - Henry Draper}, {\em HR - Harvard Obs., Revised Catalogue},  
{\em LTT - Luyten Two Tenths}, {\em  SAO - Smithsonian Astrophysical Observatory} and {\em  PPM - Position and Proper Motion}.}
\tablenotetext{c}{Coordinate data are taken from sources in the {\sc 2mass} All-Sky Release Point 
Source catalog (March 2003). [http://irsa.ipac.caltech.edu/cgi-bin/Gator/nph-dd]}
\tablenotetext{d}{Comments indicate observing conditions: ND = Neutral Density filter used - 
{\em e.g.,} 2.5 ND indicates that a 2.5-magnitude neutral density filter was inserted into 
the beam.}
\label{OBStable}
\end{deluxetable}

\clearpage

\begin{deluxetable}{llllrrrrrl}
\tablewidth{0pt}
\tabletypesize{\scriptsize}
\tablecaption{Spectral types, optical {\sc ubvri} magnitudes and special-case commentary for stars 
of known spectral type and luminosity observed during this program.}
\tablehead{
\colhead{Star\tablenotemark{a}}    & \colhead{Comments\tablenotemark{b}} & \colhead{Spectral}  
& \colhead{Ref~~~~}                & \colhead{V} & \colhead{(U-B)} & \colhead{(B-V)} & \colhead{(V-R)c} 
& \colhead{(R-I)c} & \colhead{Ref} \\
\colhead{}                         & \colhead{}         & 
\colhead{Types}                    & \colhead{SpTy}     & 
\colhead{[mags]} & \colhead{[mags]} & \colhead{[mags]} & \colhead{[mags]} & \colhead{[mags]} & 
\colhead{Photometry}}
\startdata
 HR 5190         &                  & B2 IV     & H69   &  3.41~  & -0.85  & -0.22  & -0.12  & -0.21  & JMIW66 \\
 HD 192273       &                  & B2 V      & HC75  &  8.845~ & -0.817 & -0.187 & -0.091 & -0.111 & MCBL89 \\
 HR 4638         &                  & B3 V      & H78   &  3.96~  & -0.61  & -0.15  & -0.071 & -0.094 & JMIW66, C80a \\
 C 112           &                  & B5 IIIv   & C58   &  4.36~  & -0.44  & -0.09  & -0.016 & -0.04  & JMIW66, C80a \\
 HR 5453         &                  & B5 V      & H69   &  4.05~  & -0.56  & -0.15  & -0.067 & -0.094 & JMIW66, C80a \\
 C 120           &                  & B5 V      & B57   &  4.40~  & -0.55  & -0.14  & -0.06  & -0.076 & JMIW66, C80a \\
 HR 8628         &                  & B8 Ve     & H82   &  4.16~  & -0.34  & -0.12  & -0.028 & -0.04  & JMIW66, E78 \\
 E 502           &                  & B8/9 V    & H78   &  8.061~ & -0.238 & -0.032 & -0.014 & -0.017 & MCBL89 \\
\enddata
\tablenotetext{a}{ Same primary identifiers as listed in Table~\ref{OBStable}}
\tablenotetext{b}{ESPs = Extra Solar Planet host star; SB/SB2 = Spectroscopic Binary star; 
X-ray = X-ray active star, detected in the {\sc rosat} All-Sky Survey [{\sc rass}]; 
double/multiple/triple = multiple star system, in most cases visual binaries/triples, 
as detailed by the {\sc simbad} database; HPM = High Proper Motion stars; BD-binary = 
Brown Dwarf binary also present in the system.}
\tablerefs{H78 - Houk 1978; G06 - Gray 2006; L72 - Levato 1972; M70 - McClure 1970; 
E63 - Eggen 1963; MK73 - Morgan \& Keenan 1973; HC75 - Houk \& Cowley 1975; H82 - 
Houk 1982; H69 - Hiltner, Garrison \& Schild 1969; K10 - Koen et al. 2010; 
F09 - Frasca et al. 2009; B57 - Bonsack \& Stock 1957; C58 - Crawford 1958; 
JM53 - Johnson \& Morgan 1953; BD64 - Buscombe \& Dickens 1964; 
H88 - Houk \& Smith-Moore 1988; G89 - Garcia 1989; M01 - Montes et al. 2001; 
KM89 - Keenan \& McNeil 1989; S55 - Slettebak 1955; R95 - Reid, Hawley \& Gizis 1995; 
R04 - Reid et al. 2004; G01 - Gray, Napier \& Winkler 2001; M75 - Malaroda 1975; 
S70 - Schild 1970; KY88 - Keenan \& Yorka 1988; JMIW66 - Johnson et al. 1966; 
MCBL89 - Menzies et al. 1989; M98 - Metanomski et al. 1998; C80a - Cousins 1980a; 
E78 - Eggen 1978; C83 - Cousins 1983; C80b - Cousins 1980; C80c - Cousins 1980; 
HO87 - H\"{a}ggkvist \& Oja, 1987; F66 - Feinstein 1966; R69 - Rybka 1969; 
CHW67 - Cowley, Hiltner \& Witt 1967; C73 - Cousins 1973; F83 - Fernie 1983; 
R55 - Roman 1955; C64 - Cousins 1964; J65 - Johnson 1965; 
DK84 - Dachs \& Kaiser 1984; L09 - Landolt 2009}
\label{SpTytable}
\end{deluxetable}

\clearpage

\begin{deluxetable}{lrrrc}
\tablewidth{0pt}
\tabletypesize{\scriptsize}
\tablecaption{Infrared {\sc 2mass} magnitudes for our spectral library stars.}
\tablehead{
\colhead{Star\tablenotemark{a}}    & \colhead{J} & \colhead{H} & \colhead{K} & \colhead{Err}   \\
\colhead{}                         & \colhead{[2MASS]\tablenotemark{b}} &  
\colhead{[2MASS]\tablenotemark{b}} & \colhead{[2MASS]\tablenotemark{b}} & 
\colhead{Code\tablenotemark{b}}}
\startdata
 HR 5190        &  4.014 $\pm$ 0.260 &  4.139 $\pm$  0.222 &  4.240 $\pm$ 0.288 & DDD   \\
 HD 192273      &  9.267 $\pm$ 0.022 &  9.340 $\pm$  0.023 &  9.390 $\pm$ 0.025 & AAA   \\
 HR 4638        &  4.203 $\pm$ 0.326 &  4.302 $\pm$  0.322 &  4.420 $\pm$ 0.020 & DDA   \\
 C 112          &  4.444 $\pm$ 0.284 &  4.422 $\pm$  0.244 &  4.482 $\pm$ 0.018 & DDA   \\
 HR 5453        &  4.346 $\pm$ 0.268 &  4.573 $\pm$  0.076 &  4.507 $\pm$ 0.036 & DEE   \\
 C 120          &  4.704 $\pm$ 0.037 &  4.758 $\pm$  0.029 &  4.758 $\pm$ 0.021 & EEA   \\
 HR 8628        &  4.281 $\pm$ 0.292 &  4.365 $\pm$  0.246 &  4.400 $\pm$ 0.036 & DDA   \\
 E 502          &  8.091 $\pm$ 0.025 &  8.153 $\pm$  0.034 &  8.150 $\pm$ 0.022 & AAA   \\
\enddata
\tablenotetext{a}{Same primary identifiers as listed in Table~\ref{OBStable}}
\tablenotetext{b}{Data are taken from sources in the {\sc 2mass} All-Sky Release Point 
Source catalogue (March 2003). [http://irsa.ipac.caltech.edu/cgi-bin/Gator/nph-dd]}
\label{2MASStable}
\end{deluxetable}

\clearpage

\begin{deluxetable}{llllccl}
\tablewidth{0pt}
\tabletypesize{\scriptsize}
\tablecaption{Results of spectral type homogeneity test by cross correlation.}
\tablehead{
\colhead{Target\tablenotemark{a}} & \colhead{SpTy\tablenotemark{a}} & 
\colhead{Template\tablenotemark{a}} & \colhead{SpTy\tablenotemark{a}}  
& \colhead{CCF\tablenotemark{a}} & \colhead{$\Delta$SpTy\tablenotemark{b}} 
& \colhead{$\Delta$Lumin.\tablenotemark{b}} \\
                 &                &                    &                 
& \colhead{Height}               &                        & \colhead{Class} }
\startdata
HR 5190      &  B2 IV       &   HR 5453   &   B5 V       &   0.94  &     3 &   IV $\rightarrow$ V \\
HD 192273    &  B2 V        &   HR 5190   &   B2 IV      &   0.94  &     0 &    V $\rightarrow$ IV \\
HR 4638      &  B3 V        &   C 120     &   B5 V       &   0.94  &     2 & \\
C 112        &  B5 IIIv     &   E 682     &   B9 V       &   0.97  &     4 &  IIIv $\rightarrow$ V \\ 
HR 5453      &  B5 V        &   C 120     &   B5 V       &   0.97  &     0 & \\
C 120        &  B5 V        &   HR 5453   &   B5 V       &   0.97  &     0 & \\
HR 8628      &  B8 Ve       &   E 682     &   B9 V       &   0.96  &     1 & \\
E 502        &  B8/9 V      &   E 682     &   B9 V       &   0.98  &   0.5 & \\
\enddata
\tablenotetext{a}{For each star in the spectral library (see Table~\ref{SpTytable}), 
we cross correlate against every other star in the library, and note the spectral 
match that yields the highest cross correlation function height (column 5).}
\tablenotetext{b}{Numerical difference in spectral type is noted, as well as any 
change in luminosity class between target and spectral-match template.}
\label{SAAO-selfCCF}
\end{deluxetable}

\clearpage

\begin{deluxetable}{llrrr}
\tablewidth{0pt}
\tabletypesize{\scriptsize}
\tablecaption{{\sc hipparcos} astrometric data for our spectral library stars.}
\tablehead{
\colhead{Target} & \colhead{HIP\tablenotemark{a} \#\#} & \colhead{Parallax\tablenotemark{b}} & \colhead{PMRA\tablenotemark{b}} & \colhead{PMDEC\tablenotemark{b}} \\
\colhead{}       & \colhead{}                 & \colhead{[mas]}      & \colhead{[mas/yr]} & \colhead{[mas/yr]}}
\startdata
HR 5190    & HIP  67464 & $   7.47 \pm  0.17 $ & $  -26.77 \pm  0.12 $ & $  -20.18 \pm  0.08 $ \\
HD 192273  & HIP 100170 & $   0.24 \pm  1.09 $ & $    8.78 \pm  0.70 $ & $    2.74 \pm  0.97 $ \\
HR 4638    & HIP  59449 & $   8.61 \pm  0.76 $ & $  -34.92 \pm  0.52 $ & $  -16.81 \pm  0.47 $ \\
C 112      & HIP  96468 & $   8.34 \pm  0.79 $ & $   -0.87 \pm  0.79 $ & $  -20.39 \pm  0.37 $ \\
HR 5453    & HIP  71536 & $  10.32 \pm  0.16 $ & $  -28.26 \pm  0.12 $ & $  -28.82 \pm  0.13 $ \\
C 120      & HIP 115033 & $   8.12 \pm  0.38 $ & $   17.13 \pm  0.56 $ & $  -11.70 \pm  0.40 $ \\
HR 8628    & HIP 111954 & $   6.70 \pm  0.66 $ & $   23.22 \pm  0.70 $ & $   -0.16 \pm  0.46 $ \\
E 502      & HIP  59218 & $   2.47 \pm  0.65 $ & $  -14.47 \pm  0.43 $ & $   -4.87 \pm  0.36 $ \\
\enddata
\tablenotetext{a}{{\sc hipparcos} identifiers, and astrometric data, are taken from the van Leeuwen (2007) 
New Reduction catalog available online: {\em http://vizier.u-strasbg.fr/viz-bin/VizieR?-source=I\%2F311}}
\tablenotetext{b}{Units of parallax are milli-arcsec [mas], while for proper motion in RA [PMRA] and 
proper motion in DEC [PMDEC] are both milli-arcsecs per year [mas/yr].}
\label{HIPtable}
\end{deluxetable}

\clearpage

\begin{deluxetable}{llcrccrrrc}
\tablewidth{0pt}
\tabletypesize{\scriptsize}
\rotate
\tablecaption{Properties of M7 stars observed during this program}
\tablehead{
\colhead{M7\tablenotemark{a,b,c}} & \colhead{JJ97\tablenotemark{e}}  & 
\colhead{~RA(2000)  ~  DEC(2000)} & \colhead{V\tablenotemark{g}} 
& \colhead{B-V\tablenotemark{g}}  & \colhead{V-Ic\tablenotemark{g}}  
& \colhead{J} & \colhead{H} & \colhead{K} & \colhead{Err}\\ 
\colhead{Target}  & \colhead{}        & \colhead{[2MASS]\tablenotemark{f}} & 
\colhead{(P95)} & \colhead{(P95)} & \colhead{(P95)} & \colhead{[2MASS]\tablenotemark{f}}   
& \colhead{[2MASS]\tablenotemark{f}}   & \colhead{[2MASS]\tablenotemark{f}}   & \colhead{[Code]\tablenotemark{f}}}
\startdata
    R1                    & JJ 40     &  17 51 35.20 -34 45 58.68  & 13.30  & 99.9  & 1.18  & 11.339 $\pm$ 0.030 & 10.804 $\pm$ 0.035 & 10.635 $\pm$ 0.027 & AAA  \\
    R3b                   &           &  17 51 53.07 -34 53 14.29  & 15.13  & 99.9  & 1.76  & 12.199 $\pm$ 0.042 & 11.516 $\pm$ 0.038 & 11.341 $\pm$ 0.037 & AAA  \\
    R4                    &           &  17 51 51.83 -34 50 10.24  &  9.96  & 0.24  & 0.41  &  9.298 $\pm$ 0.020 &  9.143 $\pm$ 0.024 &  9.116 $\pm$ 0.021 & AAA  \\
    R6                    &           &  17 51 52.44 -34 56 41.01  & 10.64  & 0.54  & 0.63  &  9.577 $\pm$ 0.023 &  9.369 $\pm$ 0.024 &  9.270 $\pm$ 0.023 & AAA  \\
    R7b                   & JJ 14     &  17 52 26.24 -34 35 40.85  & 13.78  & 1.08  & 1.42  & 11.461 $\pm$ 0.057 & 10.859 $\pm$ 0.065 & 10.723 $\pm$ 0.045 & AAA  \\
   R10                    &           &  17 52 32.36 -34 55 50.75  & 12.99  & 1.10  & 1.29  & 10.650 $\pm$ 0.029 & 10.068 $\pm$ 0.030 &  9.873 $\pm$ 0.026 & AAA  \\
   R11                    &           &  17 52 36.61 -34 44 39.70  & 14.23  & 1.04  & 1.31  &  9.030 $\pm$ 0.030 &  8.382 $\pm$ 0.036 &  8.231 $\pm$ 0.026 & AAA  \\
   R14                    & JJ 26     &  17 52 46.73 -34 40 58.78  & 12.06  & 0.66  & 0.77  & 10.672 $\pm$ 0.050 & 10.309 $\pm$ 0.061 & 10.195 $\pm$ 0.057 & AAA  \\
\enddata
\tablenotetext{a}{M7 target stars are drawn from Prosser et al. (1995 - P95) [R$\#$] 
and Koelbloed (1959 - K59) [K$\#$] studies.} 
\tablenotetext{b}{Koelbloed (1959) cross identifications with P95 sources: 
R4=K13; R6=K15; R16b=K45; R20=K44; R50=K61; R58a=K70; R120=K116;}
\tablenotetext{c}{High S/N ($>$100), high resolution (R$\simeq$80,000) optical spectra 
are available for several M7 stars in the {\sc uves} Paranal Observatory Project [{\sc pop}] 
library (http://www.eso.org/sci/observing/tools/uvespop.html.html - Bagnulo et al., 2003).}
\tablenotetext{d}{Indicates that K$\#$ star V/B-V photometry data are 
photographic. All other Koelbloed (1959) stars have photoelectric data.}
\tablenotetext{e}{Cross identification matches with the James \& Jeffries (1997 - JJ97) 
catalog.}
\tablenotetext{f}{Coordinate and photometric data are taken from sources in 
the {\sc 2mass} All-Sky Release Point Source catalog (March 2003). 
[http://irsa.ipac.caltech.edu/cgi-bin/Gator/nph-dd]}
\tablenotetext{g}{All V, B-V and V-Ic data are taken from P95, except for the 
K$\#$ stars whose data are taken from K59.}
\label{M7table1}
\end{deluxetable}

\clearpage

\begin{deluxetable}{lccccccccccccc}
\tablewidth{0pt}
\tabletypesize{\scriptsize}
\rotate
\tablecaption{Spectral types, photometry and reddening vectors of M7 stars observed during this program}
\tablehead{
\colhead{M7\tablenotemark{a}} & \colhead{Spectral\tablenotemark{b}}  
& \colhead{B-V}  & \colhead{K95\tablenotemark{c}}  & \colhead{K95\tablenotemark{d}} 
& \colhead{V-Ic}  & \colhead{K95\tablenotemark{c}}  & \colhead{K95\tablenotemark{d}} 
& \colhead{J-H}  & \colhead{K95\tablenotemark{c,e}}  & \colhead{K95\tablenotemark{d,e}} 
& \colhead{H-K}  & \colhead{K95\tablenotemark{c,e}}  & \colhead{K95\tablenotemark{d,e}} \\ 
\colhead{Target}  & \colhead{Type}         
& \colhead{} & \colhead{(B-V)} & \colhead{E$_{(B-V)}$} 
& \colhead{} & \colhead{(V-Ic)} & \colhead{E$_{(V-Ic)}$} 
& \colhead{} & \colhead{(J-H)} & \colhead{E$_{(J-H)}$} 
& \colhead{} & \colhead{(H-K)} & \colhead{E$_{(H-K)}$} }
\startdata
%
 R1      & K3V      &  99.9 &  0.97 &  99.9  &   1.18  &  1.08 &   0.10  &  0.535 &   0.484 &  0.051 &  0.169 &  0.118 &  0.051 \\
 R3b     & K5/7e    &  99.9 &  1.16 &  99.9  &   1.76  &  1.36 &   0.40  &  0.683 &   0.553 &  0.130 &  0.175 &  0.138 &  0.037 \\
 R4      & A7V      &  0.24 &  0.19 &   0.05 &   0.41  &  0.27 &   0.14  &  0.155 &   0.033 &  0.122 &  0.027 &  0.058 & -0.031 \\
 R6      & F9V      &  0.54 &  0.55 &  -0.01 &   0.63  &  0.69 &  -0.06  &  0.208 &   0.249 & -0.041 &  0.099 &  0.078 &  0.021 \\
 R7b     & K5Ve     &  1.08 &  1.16 &  -0.08 &   1.42  &  1.36 &   0.06  &  0.602 &   0.553 &  0.049 &  0.136 &  0.138 & -0.002 \\
 R10     & K4Ve     &  1.10 &  1.07 &   0.03 &   1.29  &  1.15 &   0.14  &  0.582 &   0.523 &  0.059 &  0.195 &  0.128 &  0.067 \\
 R11     & K4V      &  1.04 &  1.07 &  -0.03 &   1.31  &  1.15 &   0.16  &  0.648 &   0.523 &  0.125 &  0.151 &  0.128 &  0.023 \\
 R14     & G1V      &  0.66 &  0.60 &   0.06 &   0.77  &  0.72 &   0.05  &  0.363 &   0.259 &  0.104 &  0.114 &  0.078 &  0.036 \\
\enddata
\tablenotetext{a}{M7 target stars, as identified in Table~\ref{M7table1}}
\tablenotetext{b}{Spectral type determinations from this study.}
\tablenotetext{c}{Theoretical B-V, V-Ic, J-H and H-K colors for each 
target's spectral type (assumes dwarf class) are taken from Kenyon \& 
Hartmann (1995 - KH95).}
\tablenotetext{d}{E$_{(B-V)}$, E$_{(V-Ic)}$, E$_{(J-H)}$, E$_{(H-K)}$ vectors 
are calculated by subtracting each target's respective color from its 
theoretical KH95 counterpart (whose values are based on spectral type).}
\tablenotetext{e}{KH95 JHK magnitudes, based on the Bessel \& Brett (1988) 
system, have been transformed onto the {\sc 2mass} system using equations 
presented in Carpenter (2001).}
\label{M7table2}
\end{deluxetable}


\begin{thebibliography}{}
\bibitem[Abbott et al. (2005)]{DES} Abbott, T., et al. 2005, arXiv:astro-ph/0510346
\bibitem[Abdalla et al. (2012)]{DESpec} Abdalla,~ F., et al. 2012, 
arXiv:1209.2451 [astro-ph.CO]
\bibitem[Abt (1975)]{abt75} Abt,~ H.~A., 1975, \pasp, 87, 417
\bibitem[Allende Prieto (2004)]{S4N2004} Allende Prieto,~ C., Barklem,~ P.S., 
Lambert,~ D.~L., \& Cunha,~ K., 2004, \aap, 420, 183
\bibitem[Bagnulo et al. (2003)]{uvesPOP} Bagnulo,~ S., Cabanac,~ R., Jehin,~ E., 
Ledoux,~ C., \& Melo,~ C., 2003, Messenger, 114, 10
\bibitem[Barnes (2003)]{B03} Barnes, S.~A., 2003, \apj, 586, 464
\bibitem[Barnes (2007)]{B07} Barnes, S.~A., 2007, \apj, 669, 1197
\bibitem[Barnes (2010)]{B10} Barnes, S.~A., 2010, \apj, 722, 222
\bibitem[Basri et al. (1996)]{Basri81} Basri,~ G., Marcy,~ G., \& Graham,~ J. 
1996, \apj, 458, 600
\bibitem[Bessel (1990)]{UBVRI} Bessell,~ M.S., 1990, \pasp, 102, 1181
\bibitem[Bessel \& Brett 1988]{besselJHHK} Bessel,~ M.S., \& Brett,~ J.M., 1988, \pasp, 100, 1134
\bibitem[Bochanski et al. (2006)]{SDSS-Mcat} Bochanski,~ J.~J., West,~ A.~A., 
Hawley,~ S.~L., \& Covey,~ R.~R., 2006, \aj, 133, 531
\bibitem[Bonsack \& Stock (1957)]{B57} Bonsack,~ W.~K, \& Stock,~ J., 1957, \apj, 126, 99 [B57]
\bibitem[Briel \& Pfeffermann (1986)]{pfef86} Briel,~ U.~G., \& Pfeffermann,~ E., 1986, Nuclear Instruments \& Methods in Physics Research, 242, 376
\bibitem[Buscombe \& Dickens (1964)]{BD64} Buscombe,~ W., \& Dickens,~ C.~R., 1964, \mnras, 128, 499 [BD64]
\bibitem[Cargile \& James (2010)]{IC4665-DJJPAC2010} Cargile,~ P.A., \& 
James,~ D.J., 2010, \aj, 140, 677
\bibitem[Cargile et al. (2010)]{B1-LDB} Cargile,~ P.A., James,~ D.J., \& 
Jeffries,~ R.D., 2010, \apjl, 725, L111
\bibitem[Carpenter (2001)]{2MASS-filters}  Carpenter,~ J.M., 2001, \aj, 121, 2851
\bibitem[Castander (1998)]{SDSS} Castander,~ F.J., 1998, Astrophysics and 
Space Science, 263, 91
\bibitem[Cousins (1964)]{C64} Cousins,~ A.~W.~J., 1964, Monthly Notes of the Astromonical Society of South 
Africa, 23, 175 [C64]
\bibitem[Cousins (1973)]{C73} Cousins,~ A.~W.~J., 1973, Monthly Notes of the Astromonical Society of South 
Africa, 32, 117 [C73]
\bibitem[Cousins (1980a)]{C80a} Cousins,~ A.~W.~J., 1980a, South African Astronomical Observatory Circulars, 1, 234 [C80a]
\bibitem[Cousins (1980b)]{C80b} Cousins,~ A.~W.~J., 1980b, Monthly Notes of the Astromonical Society of South 
Africa, 39, 22 [C80b]
\bibitem[Cousins (1980c)]{C80c} Cousins,~ A.~W.~J., 1980c, South African Astronomical Observatory Circulars, 1, 166 [C80c]
\bibitem[Cousins (1983)]{C83} Cousins,~ A.~W.~J., 1983, South African Astronomical Observatory Circulars, 7, 36 [C83]
\bibitem[Cowley et al. (1967)]{CHW67} Cowley,~ A.~P., Hiltner,~ W.~A., Witt,~ A.~N. 1967, \aj, 72, 1334 [CHW67]
\bibitem[Crawford (1958)]{C58} Crawford,~ D.~L., 1958, \apj, 128, 185 [C58]
\bibitem[Dachs \& Kaiser (1984)]{DK84} Dachs,~ J., \ Kaiser, D., 1985, \aaps, 58, 411 [DK84]
\bibitem[David et al. (1996)]{David96} David,~ L.~P., Harnden,~ F.~R.~J., 
Kearns,~ K.~R., and Zombeck,~ M.~V., 1996, {\sc rosat} High Resolution 
Imager Calibration Report, {\sc us rosat} Science Data Center/{\sc SAO}.
\bibitem[Demarque \& Larson (1964)]{DL64} Demarque,~ P.R., \& Larson,~ R.B., 
1964, \apj, 140, 1544
\bibitem[Doroshkevic et al. (2004)]{SDSS04} Doroshkevich,~ A., Tucker,~ D.L., 
Allam,~ S., \& Way,~ M.J., 2004, \aap, 418, 7
\bibitem[Eggen (1963)]{E63} Eggen,~ O.~J., 1963, \aj, 68, 483 [E63]
\bibitem[Eggen (1978)]{E78} Eggen,~ O.~J., 1978, \apjs, 37, 251 [E78]
\bibitem[Engberg (1983)]{engberg83} Engberg,~ M., 1983, \aaps, 54, 203
\bibitem[Feinstein (1966)]{F66} Feinstein,~ A., 1966, Information Bulletin of the Southern Hemisphere, 8, 29 [F66]
\bibitem[Fernie (1983)]{F83} Fernie,~ J.~D., 1983, \apjs, 52, 7 [F83]
\bibitem[Folkes et al. (1999)]{2dF99} Folkes,~ S., et al. 1999, \mnras, 308, 459
\bibitem[Frasca et al. (2009)]{F09}  Frasca,~ A., Covino,~ E., Spezzi,~ L., Alcala,~ J.~M., 
Marilli,~ E., Furesz,~ G., Gandolfi,~ D, 2009, \aap, 508, 1313 [F09]
\bibitem[Garcia (1989)]{G89} Garcia,~ B., 1989, Bulletin d'Information du 
Centre de Donn\'{e}es Stellaires, 36, 27 [G89]
\bibitem[Gieseking (1985)]{Giesek} Gieseking,~ F., 1985, \aaps, 61, 75
\bibitem[Gray, Napier \& Winkler (2001)]{G01} Gray,~ R.~O., Napier,~ M.~G., \& Winkler,~ L.~I., 2001, \aj, 121, 2148 [G01]
\bibitem[Gray et al. (2006)]{G06} Gray,~ R.~O., Corbally,~ C.~J., Garrison,~ R.~F., McFadden,~ M.~ T., 
Bubar,~ E.~J., McGahee,~ C.~ E., O'Donoghue,~ A.~A., \& Knox,~ E.~ R., 2006, \aj, 132, 161 [G06]
\bibitem[H\"{a}ggkvist \& Oja (1987)]{HO87} H\"{a}ggkvist,~ L., \& Oja, T., 1987, \aaps, 68, 259 [HO87]
\bibitem[Hiltner, Garrison \& Schild (1969)]{H69} Hiltner,~ W.~A., Garrison,~ R.~F., Schild,~ R.~E., 1969, 
\apj, 157, 313 [H69]
\bibitem[Houk (1978)]{H78} Houk,~ N., 1978, Michigan Spectral Survey, Ann Arbor, [H78]
Dept. of Astronomy, University of Michigan, Vol 2. [H78]
\bibitem[Houk \& Cowley (1975)]{HC75} Houk,~ N., \& Cowley,~ A.~P., 1975, Michigan Spectral Survey, Ann Arbor, 
Dept. of Astronomy, University of Michigan, Vol 1. [HC75]
\bibitem[Houk (1982)]{H82} Houk,~ N., 1982, Michigan Spectral Survey, Ann Arbor, Dept of. Astronomy, 
University of Michigan, Vol 3. [H82]
\bibitem[Houk \& Smith-Moore (1988)]{H88} Houk,~ N., \& Smith-Moore,~ M., 1988, Michigan Spectral Survey, 
Ann Arbor, Dept. of Astronomy, University of Michigan, Vol 4. [H88]
\bibitem[Ivezic et al. (2008)]{LSST} Ivezic,~ Z., et al. 2008, Serbian 
Astronomical Journal, 176, 1
\bibitem[James \& Jeffies (1997)]{JJ97} James,~ D.J., \& Jeffries,~ R.D., 1997, \mnras, 291, 252
\bibitem[James et al. (2006)]{FeH} James,~ D.J., Melo,~ C., Santos,~ N.C., 
\& Bouvier,~ J., 2006, \aap, 446, 971
\bibitem[James et al. (2010)]{M34gyro} James,~ D.J., Barnes, S.A., Meibom,~ S., 
Lockwood,~ G.W., Levine,~ S.E., Deliyannis, C., Platais,~ I., Steinhauer,~ A., 
\& Hurley,~ B.K., 2010, \aap, 515, 100
\bibitem[James et al. (2013)]{FeH3} James,~ D.~J., Aarnio,~ A.~N., Richert,~ A., Melo,~ C., 
Santos,~ N.C., \& Bouvier,~ J., 2013, \mnras, {\em in prep} 
\bibitem[Jeffries et al. (1996)]{RDJ96} Jeffries,~ R.~D., Buckley,~ D.~A.~H., 
James,~ D.~J., \& Stauffer,~ J.~R., 1996, \mnras, 281, 1001
\bibitem[Johnson \& Morgan (1953)]{JM53} Johnson,~ H.~L., \& Morgan,~ W.~W., 1953, \apj, 117, 313 [JM53]
\bibitem[Johnson (1965)]{J65} Johnson,~ H.~L., 1965, \apj, 141, 170 
\bibitem[Johnson et al. (1966)]{JMIW66} Johnson,~ H.~L., Mitchell,~ R.~I., Iriarte,~ B., \& 
Wisniewski,~ W.Z., 1966, {\em Comm. Lun. Plan. Lab. IV}, No. 63, P.99 [JMIW66]
\bibitem[Keenan \& Pitts (1980)]{KP80} Keenan,~ P.C., \& Pitts,~ R.E., 
1980, \apjs, 42, 541
\bibitem[Keenan \& Yorka (1988)]{KY88} Keenan,~ P.~C., \& Yorka,~ S.~B., 1988, Bulletin d'Information du 
Centre de Donn\'{e}es Stellaires, No. 35, P. 37
\bibitem[Keenan \& McNeil (1989)]{KM89} Keenan,~ P.C., \& McNeil,~ R.C., 1989, \apjs, 71, 245
\bibitem[Kenyon \& Hartmann (1995)]{KH95} Kenyon,~ S.~J., \& Hartmann,~ L., 1995, \apjs, 101, 117 [KH95]
\bibitem[Kilkenny \& Cousins (1995)]{Eregion2} Kilkenny,~ D., \& Cousins,~ A.W.J., 
1995, Astrophysics and Space Science, 230, 155
\bibitem[Kirkpatrick et al. (1999)]{kirk99} Kirkpatrick,~ J.D., Reid,~ I.N., 
Liebert,~ J., Cutri,~ R.M., Nelson, B., Beichman,~ C.A., Dahn,~ C.C., 
Monet,~ D.G., Gizis,~ J.E., \& Skrutskie,~ M.F., 1999, \apj, 519, 802
\bibitem[Koelbloed (1959)]{K59} Koelbloed,~ D., 1959, Bull. Astron. Inst. Netherands 14, 265 [K59]
\bibitem[Koen et al. (2010)]{K10} Koen,~ C., Kilkenny,~ D., van Wyk,~ F., \& Marang,~ F., 2010, \mnras, 403, 1949 [K10]
\bibitem[Landolt (2009)]{L09} Landolt,~ A.~U., 2009, \aj, 137, 4186 [L09]
\bibitem[Lebzelter et al. (2012)]{CRIRES-POP} Lebzelter,~ T., et al., 2012, 
\aap, 539, 109
\bibitem[Leung \& Schneider (1975)]{leung75} Leung,~ K.~C., \& Schneider,~ D.~P., 1975, \apj, 201, 792
\bibitem[Levato (1972)]{L72} Levato,~ H., \pasp, 84, 584 [L72]
\bibitem[Maitzen \& Floquet (1981)]{MFloq81} Maitzen,~ H.~M., \& Floquet,~ M.,~ 1981, \aap, 100, 3 
\bibitem[Malaroda (1975)]{M75} Malaroda,~ S., 1975, \aj, 80, 637 [M75]
\bibitem[Mart\'{\i}n et al. (1998)]{Martin98} Mart\'{\i}n,~ E.L., 
Montmerle,~ T., Gregorio-Hatem,~ J., \& Casanova,~ S., 1998, \mnras, 300, 733
\bibitem[McClure (1970)]{M70} McClure,~ R.~D., 1970, \aj, 75, 41 [M70]
\bibitem[Menzies \& Laing (1988)]{F-region} Menzies,~ J.~W., \& Laing,~ J.~D., 
1998, \mnras, 231, 1047
\bibitem[Menzies et al. (1989)]{MCBL89} Menzies,~ J.~W., Cousins,~ A.~W.~J., Banfield,~ R.~ M.,  
\& Laing,~ J.~D., 1989, South African Astronomical Observatory Circulars, 13, 1 [MCBL89]
\bibitem[Mermilliod (1981)]{merm81} Mermilliod,~ J.-C., 1981, \aap, 97, 235 
\bibitem[Metanomski et al. (1998)]{M98} Metanomski,~ A.~D.~F., Pasquini,~ L., Krautter,~ J., 
Cutispoto,~ G., \& Fleming,~ T.~A., 1998, \aaps, 131, 197
\bibitem[Meynet et al. (1993)]{meynet} Meynet,~ G., Mermilliod,~ J.-C., \& Maeder,~ A., 
1993, \aaps, 98, 477
\bibitem[Minniti et al. (2006)]{VVV} Minniti,~ P., et al. 2006, 
{\em Memorie della Societ\`{a} Astronomica Italiana}, v77, p.1184
\bibitem[Montes et al. (1997)]{Montes97} Montes,~ D., Mart\'{\i}n,~ E.L., 
Fern\'{a}ndez-Figueroa,~ M.J., Cornie,~ M., \& De Castro,~ E., 1997, 
\aaps, 123, 473
\bibitem[Montes et al. (2001)]{M01} Montes,~ D., Lopez-Santiago,~ J., Galvez,~ M.~C., 
Fernandez-Figueroa,~ M.~J., De Castro,~ E., \& Cornide,~ M., 2001, \mnras, 328, 45 [M01]
\bibitem[Morgan \& Keenan (1973)]{MK73} Morgan,~ W.~W., \& Keenan,~ P.~C., 1973, 
Annual Review of Astronomy and Astrophysics, 11, 29 [MK73]
\bibitem[Naylor (2009)]{Naylor09} Naylor,~ T., 2009, \mnras, 399, 432
\bibitem[Prosser et al. (1995)]{P95} Prosser,~ C.~F., Stauffer,~ J.~R., 
Caillault,~ J-P., Balachandran,~ S., Stern,~ R.~A., \& Randich,~ S., 
1995, \aj, 110, 1229
\bibitem[Ram{\'{\i}}rez \& Mel\'{e}ndez (2005)]{RAM2005} Ram{\'{\i}}rez,~ I., 
\& Mel\'{e}ndez,~ J., 2005, \apj, 626, 465
\bibitem[Reid, Hawley \& Gizis (1995)]{R95} Reid,~ I.~N., Hawley,~ S.~L., Gizis,~ J.~E., 1995, \aj, 110, 1838 [R95]
\bibitem[Reid et al. (2004)]{R04} Reid,~ I.~N., Cruz,~ K.~L., Allen,~ P., et al. 2004, \aj, 128, 463 [R04]
\bibitem[Roman (1955)]{R55} Roman,~ N.~G., 1955, \apjs, 2, 195
\bibitem[Rybka (1969)]{R69} Rybka,~ E., 1969, Acta Astronautica, 19, 229 [R69]
\bibitem[Sandage (1958)]{sandage58} Sandage,~ A., 1958, Ricerche Astronomiche, 
Vol. 5, Specola Vaticana, Proceedings of a Conference at Vatican Observatory, 
Castel Gandolfo, May 20-28, 1957, Amsterdam: North-Holland, and New York: 
Interscience, 1958, edited by D.J.K. O'Connell., p.41
\bibitem[Schild (1970)]{S70} Schild,~ R.~E., 1970, \apj, 161, 855 [S70]
\bibitem[Sestito et al. (2003)]{M7sestito} Sestito,~ P., Randich,~ S., 
Mermilliod,~ J.-C., \& Pallavicini,~ R., 2003, \aap, 407, 289
\bibitem[Slettebak (1955)]{S55} Slettebak,~ A., 1955, \apj, 121, 653 [S55]
\bibitem[Snowden (1976)]{M7snowden} Snowden,~ M.~S., 1976, \pasp, 88, 174
\bibitem[Spencer Jones (1980)]{SAAOextin} Spencer Jones, J.~H., 1980, Monthly 
Notes of the Astronomical Society of Southern Africa (MNSSA), 39, 89
\bibitem[Stone \& Baldwin (1983)]{LTT7379} Stone,~ R.~P.~S., \& Baldwin,~ J.~A., 1983, \mnras, 204, 347
\bibitem[Tonry \& Davis (1979)]{TD79} Tonry,~ J., \& Davis,~ M., 1979, \aj, 84, 1511
\bibitem[van Leeuwen (2007)]{newHIP} van Leeuwen,~ F., 2007, A\&A, 474, 653
\bibitem[van Leeuwen (2009)]{HIPclusters} van Leeuwen,~ F., 2009, A\&A, 497, 209
\bibitem[Villanova et al. (2009)]{M7-villa} Villanova,~ S., Carraro,~ G., \& Saviane,~ I., 2009, \aap, 504, 845
\bibitem[Vogt et al. (1981)]{E-region1} Vogt,~ N., Geisse,~ H.S., \& Rojas,~ S., 1981, \aaps, 46, 7


\end{thebibliography}
\end{document}